\documentclass[conference,10pt,a4paper]{IEEEtran}
\IEEEoverridecommandlockouts
\usepackage{cite}
\usepackage{amsmath,amssymb,amsfonts,bm}
\usepackage{algorithm}
\usepackage{algorithmic}
\usepackage{graphicx}
\usepackage{xcolor}
\usepackage{comment}

\usepackage[acronyms]{glossaries}

\usepackage{pgfplots}
\pgfplotsset{compat=newest}
\usepgfplotslibrary{groupplots} 
\usepackage{tikz}
\usetikzlibrary{calc}
\usetikzlibrary{positioning}
\usetikzlibrary{arrows.meta}
\usetikzlibrary{spy}

\usetikzlibrary{external}
\usepackage[hidelinks]{hyperref}

\newcommand{\transpose}{\text{T}}
\newcommand{\hermitian}{\text{H}}
\newcommand{\diag}{\text{diag}} 
\newcommand{\trace}{\text{tr}}

\newcommand{\imagUnit}{i}

\newcommand{\ist}{\hspace*{.3mm}}
\newcommand{\rmv}{\hspace*{-.3mm}}
\newcommand{\iist}{\hspace*{1mm}}

\newcommand{\nn}{\nonumber}

\newacronym{sbl}{SBL}{sparse Bayesian learning}
\newacronym{nomp}{NOMP}{newtonized orthogonal matching pursuit}
\newacronym{ml}{ML}{maximum likelihood}
\newacronym{pdf}{PDF}{probability density function}
\newacronym{snr}{SNR}{signal-to-noise ratio}
\newacronym{em}{EM}{expectation-maximization}
\newacronym{awgn}{AWGN}{additive white Gaussian noise}
\newacronym{doa}{DOA}{direction of arrival}
\newacronym{mmv}{MMV}{multiple measurement vector}
\newacronym{ospa}{OSPA}{optimal subpattern assignment}
\newacronym{mimo}{MIMO}{multiple-input multiple-output}
\newacronym{prf}{PRF}{pulse reprtition frequency}
\newacronym{fov}{FOV}{field of view}
\newacronym{cfar}{CFAR}{constant false alarm rate}
\newacronym{tbd}{TBD}{track-before-detect}
\newacronym{mot}{MOT}{multiobject tracking}
\newacronym{lasso}{LASSO}{least absolute shrinkage and selection operator}
\newacronym{music}{MUSIC}{multiple signal classification}
\newacronym{esprit}{ESPRIT}{estimation of signal parameters via rotational invariant techniques}
\newacronym{mp}{MP}{matching pursuit}

\colorlet{plot1}{cyan!30!white}
\colorlet{plot2}{red!40!white}
\colorlet{plot3}{cyan!70!black}
\colorlet{plot4}{black}

\begin{document}
	\title{\huge A Block-Sparse Bayesian Learning Algorithm with Dictionary Parameter Estimation for Multi-Sensor Data Fusion \\ 
		\thanks{This project was partly funded by the Christian
Doppler Laboratory for Location-aware Electronic Systems.}
	}
	
	\author{
	\IEEEauthorblockN{Jakob M\"oderl$^\ast$, Anders Malte Westerkam$^{\dagger}$, Alexander Venus$^\ast$, and Erik Leitinger$^\ast$}
    \IEEEauthorblockA{$^*$Graz University of Technology, Graz, Austria,  \{jakob.moederl, a.venus, erik.leitinger\}@tugraz.at}
    \IEEEauthorblockA{$^\dagger$Aalborg University, Aalborg Denmark, amw@es.aau.dk}
    }
	\maketitle
	
	\begin{abstract}
         We propose an \gls{sbl}-based method that leverages group sparsity and multiple parameterized dictionaries to detect the relevant dictionary entries and estimate their continuous parameters by combining data from multiple independent sensors. In a MIMO multi-radar setup, we demonstrate its effectiveness in jointly detecting and localizing multiple objects, while also emphasizing its broader applicability to various signal processing tasks. A key benefit of the proposed \gls{sbl}-based method is its ability to resolve correlated dictionary entries---such as closely spaced objects---resulting in uncorrelated estimates that improve subsequent estimation stages.

        Through numerical simulations, we show that our method outperforms the \gls{nomp} algorithm when two objects cross paths using a single radar. Furthermore, we illustrate how fusing measurements from multiple independent radars leads to enhanced detection and localization performance.
    \end{abstract}
	

    \glsresetall
	
	\section{Introduction}    
	\label{sec:introduction}

    Accurate detection, localization, and tracking of objects is the cornerstone of numerous real-world applications, including ocean science \cite{JanMeySnyWigBauHil:JASA}, integrated sensing and communications \cite{Nuria:ProcIEEE2024}, and autonomous robotics \cite{LevinsonIV2011}. Despite their fundamental importance, these tasks become particularly challenging in scenarios characterized by highly cluttered environments and multiple closely spaced objects.
    In radar and related applications, the received signal can be modeled as linear combination of the radar's response to each individual object in the scene.
    Furthermore, the radar's response to each object is parameterized by the objects location - the parameter of interest.
    Jointly estimating (i) then number of components in a superimposed mixture and (ii) the parameters of each component in the mixture is fundamental for detecting and tracking objects, as well as for other related signal processing applications.
    Jointly addressing (i) and (ii) while fusing multiple observations---such as data from several independent radars monitoring the same area---introduces additional challenges but can also yield significant performance gains, for example, by leveraging the different vantage points of the sensors.

    Conventional radar processing chains apply a detect-then-track approach which typically relies on matched filtering in time and space and cell-based detectors, such as the \gls{cfar} detector, for preprocessing the raw radar signals \cite{Niz:TAES1972,MalZhi:TSP1993}.
    The detections of this preprocessing stage are then used as inputs to \gls{mot} methods, e.g., based on Bayesian inference \cite{BarWilTia:B11, Mahler2007, MeyerProc2018, LiLeiVenTuf:TWC2022,VenLeiTerMeyWit:TWC2024}.
    Such two-step approaches significantly reduce data flow and are computationally efficient. Therefore, they are widely used in practice.
    Nevertheless, two-step approaches have notable limitations, e.g., when faced with low \gls{snr} objects and closely spaced objects whose signals become unresolvable after beamforming.
    One possible way to address this shortcoming is to use \gls{tbd} methods, which operate directly on matched-filtered radar signals \cite{DavWieVu:JSTSP2013, PapVoVoFanBea:TAES2015 ,Ristic2020, KroWilMey:FUSION2021,Kropfreiter2024} or even on raw signals \cite{westerkamFusion2025,LiaLeiMey:Asilomar2023,LiaLeiMey:TSP2024,LiaMey:Asilomar2024}, thereby enhancing tracking performance in weak-object scenarios and improving the resolution of closely spaced objects. However, these benefits come at a considerable increase in computational complexity.

    We apply an alternate approach to \gls{tbd} by decreasing the information loss of the preprocessing stage to improve tracking performance.
    To overcome the limitations of a cell-based detection stage, classic parameter estimation methods such as \gls{music} and \gls{esprit} \cite{stoica2005:SpectralAnalysis}, assume that the number of objects (i.e., the model order) is known, which is typically not the case for radar and similar applications.
    Sparse signal reconstruction methods, such as the \gls{lasso} \cite{tibshirani1996RSSB}, \gls{mp} \cite{mallat1993TSP:MP} or \gls{sbl} \cite{Faul2001,palmerNIPS2003}, solve the problem of estimating the number of objects in an efficient way by modeling the observed signal as a product of a large dictionary matrix with a sparse amplitude vector.
    Thus, the number of objects is indirectly estimated as the number of nonzero amplitudes.
    Note that many sparse reconstruction methods can be unified within a common framework \cite{wipf2011TIP}.
    In \cite{wipf2011TIP}, \gls{lasso} and \gls{mp} are classified as Type-I Bayesian methods, whereas \gls{sbl} belongs to the more general and typically superior Type-II methods. Furthermore, \gls{sbl} is closely linked to stepwise regression \cite{AmeGomICML2021:SBLStewpwiseRegression} and can be implemented with a computational complexity comparable to that of \gls{mp} \cite{PotRaoICASSP2023:lightweight-SBL}.
    
    While initially developed for fixed dictionaries, sparse signal reconstruction methods can be extended to estimate the parameters (e.g., the location of objects) on a continuum by considering a parameterized dictionary matrix \cite{hansen2014SAM:SBL, GreLeiWitFle:TWC2024, hansen2018TSP:SuperFastLSE}, which further helps to improve the performance for correlated (e.g., closely-spaced) objects \cite{duarte2013ACHA, chi2011TSP}.
    Multiple radar-sensors observing the same scene, i.e., the same objects, result in the amplitude vectors corresponding to each sensor/observation to share the same sparsity-pattern. Thus, resulting in a group-sparse signal reconstruction problem. Naturally, sparse signal reconstruction methods have been extended to group-sparsity, e.g., group-\gls{lasso} \cite{yuan2006RSSB}, block-\gls{mp} \cite{eldar2010TSP:BlockSparseMP} or block-\gls{sbl} \cite{zhang2013TSP:BlockSparseSBL, luessi2013TSP:VariationalBSBL, babacan2014TSP:variationalBSBL, NanGemGerHidMec:SP2019, MoePerWitLei:Arxiv2023BSBL, MoePerWitLei:TSP2024}. However, none of these methods estimate the dictionary parameters on a continuum, except for \cite{MoePerWitLei:TSP2024} which focuses on group-sparsity within a single observation vector and unknown group-sizes rather than multiple independent observations.

    In this work, we propose a novel preprocessing method for sensor fusion between multiple independent sensors (e.g., mutually independent radars) based on \gls{sbl} to jointly detect relevant dictionary entries and estimate their parameters on the continuum.
    The proposed \gls{sbl}-based method extends the block-\gls{sbl} method introduced in \cite{MoePerWitLei:Arxiv2023BSBL} to multiple parameterized dictionary matrices, making it applicable to  multi-sensor data fusion. 
    For \gls{mot} based on data from multiple radars, our method significantly enhances the measurements used by Bayesian \gls{mot} algorithms \cite{BarWilTia:B11, Mahler2007, MeyerProc2018, LiLeiVenTuf:TWC2022, VenLeiTerMeyWit:TWC2024} by offering (i) more sensitive object-waveform-related detection, (ii) improved resolvability of closely spaced objects, and (iii) reduced measurement correlation for effectively “point-like” detections.
    The main contributions of this paper are as follows.
    
    \begin{itemize}
        \item We introduce an \gls{sbl}-based method that utilizes multiple parameterized dictionaries to detect relevant dictionary entries and estimates the corresponding parameters (on the continuum) by fusing data from multiple independent sensors.
        
        \item We apply the proposed algorithm to detect and localize objects in a \gls{mimo}-radar setup.

        \item For a setup consisting of a single \gls{mimo}-radar system, we demonstrate that the proposed method outperforms the \gls{nomp} algorithm \cite{mamandipoor2016TSP:NOMP} when the objects are closely spaced.
        
        \item We show that the proposed method is able to fuse the data from multiple independent \gls{mimo}-radars to enhance the detection and localization accuracy compared to the single-radar case.
    \end{itemize}

	\section{Signal model}
	\label{sec:signal-model}
	
\begin{figure}
		\centering
		\includegraphics{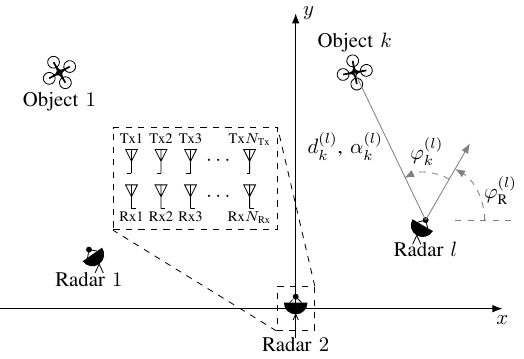}
		\caption{The scenario  considered with an unknown number of objects $K$, each located at position $\bm{\theta}_k$ being observed by $L\geq 1$ \acrshort{mimo} radars.}
		\label{fig:Overveiw}
	\end{figure}

    We consider a multi-radar setup that aims to detect and localize an unknown number of objects with some area of interest by $L \geq 1$ \gls{mimo} radars, as illustrated in Figure~\ref{fig:Overveiw}. 
    Each radar consists of $N_{\mathrm{Tx}}$ and $N_{\mathrm{Rx}}$ co-located receive and transmit antennas, respectively, that surveil the area of interest.
    Each radar is assumed to processes its own signals fully coherent, i.e., $N_{\mathrm{Tx}}\cdot N_{\mathrm{Rx}}$ channel frequency responses are obtained for each transmission from each radar.
    The signals from multiple radars are considered to be well separated in time, frequency, or code, so that multiple radars operate independently and do not observe and or interfere with each others signals.
    
    We assume that, after down-conversion to baseband, matched filtering, sampling, and transformation into the frequency domain, the received signal at the $l$th radar $(l=1,\dots,L)$ can be written as\vspace{-2mm}
    \begin{align}
        \label{eq:signal-model-sum}
        \bm{y}^{(l)} \;=\; \sum_{k=1}^{K} \bm{\psi}^{(l)}\bigl(\bm{\theta}_k\bigr)\,\alpha_k^{(l)} \;+\; \bm{v}^{(l)}\\[-7mm]\nn
    \end{align}
    where $\bm{y}^{(l)}\in \mathbb{C}^{N\times 1}$ is the backscattered signal received by the $l$th radar from $K$ objects. Here, $N_{\text{f}}$ denotes the number of samples obtained in the frequency domain for each Rx/Tx antenna pair and $N = N_{\mathrm{f}}\, N_{\mathrm{Tx}}\, N_{\mathrm{Rx}}$ denotes the total number of available samples per radar. Note that the number of objects $K$ is unknown. The complex amplitude $\alpha_k^{(l)}$ corresponds to the reflection from the $k$th object, located at $\bm{\theta}_k = [x_k \ist\iist y_k]^\transpose$, in direction of the $l$th radar.
    The function $\bm{\psi}^{(l)}(\bm{\theta})$ is the array response of the $l$th radar to an object at position $\bm{\theta} = [x \ist\iist y]^\transpose$, considering only the direct path.
    Finally, $\bm{v}^{(l)}\in\mathbb{C}^{N}$ is circularly symmetric complex AWGN with covariance $\bigl(\lambda^{(l)} \,\bm{\Lambda}_{\mathrm{v}}^{(l)}\bigr)^{-1}$, where the spectral envelope given by $\bm{\Lambda}_{\mathrm{v}}^{(l)}$ is assumed to be known (e.g., $\bm{\Lambda}_{\mathrm{v}}^{(l)}=\bm{I}$), but the total noise powers of each sensor, represented by the scalars $\lambda^{(l)}$, $l=1,\dots,L$, are treated as a nuisance parameters.

    In the following, we assume that (i) the signals are narrowband, (ii) the objects are located in the far-field of the array, and (iii) we consider a MIMO radar with $N_{\text{Tx}}=N_{\text{Rx}}=3$ transmit and receive antennas. The transmit antennas are spaced with $\lambda_{\text{c}}/2$ each whereas the receive antennas are spaced with $\lambda_{\text{c}}$, where $\lambda_{\text{c}}$ denotes the carrier wavelength, resulting in a virtual linear array with the distances between the elements and the center of the array given by $\bm{p}=\lambda_{\text{c}}\cdot [-1.5\ist\iist \rmv\rmv-\rmv\rmv 1 \ist\iist \rmv\rmv-\rmv\rmv 0.5\ist\iist \rmv\rmv-\rmv\rmv 0.5 \ist\iist 0 \ist\iist 0.5 \ist\iist 0.5 \ist\iist 1 \ist\iist 1.5]^\transpose$. The broadside direction of the array is denoted as $\varphi_{\mathrm{R}}^{(l)}$. For each Tx/Rx antenna pair we obtain a channel frequency response sampled at $N_{\text{f}}=15$ points covering a total bandwidth of $20\,\text{MHz}$.
    Given assumptions (i)--(iii) above, the response of the $l$th radar with an array centered at $\bm{\theta}_{\mathrm{R}}^{(l)} = [x_{\mathrm{R}}^{(l)} \ist\iist y_{\mathrm{R}}^{(l)}]^\transpose$ can be factorized into an angle-dependent and a distance-dependent part, i.e., $\bm{\psi}^{(l)}(\bm{\theta}) = \tfrac{1}{\sqrt{N}}\bigl(\bm{\psi}_{\varphi}(\varphi^{(l)}) \otimes \bm{\psi}_{\mathrm{d}}(d^{(l)})\bigr)$, where $\varphi^{(l)}$ and $d^{(l)}$ are functions of $\bm{\theta}$, and $\otimes$ denotes the Kronecker-product \cite{richter2005PhD}.
    Assuming isotropic antennas, the angle-dependent component is given by $\bm{\psi}_{\varphi}(\varphi^{(l)}) = e^{-\imagUnit\,2\pi\,\sin(\varphi^{(l)})\,\bm{p}/\lambda_{\text{c}}}$, where $e^{\bm{z}}$ (for a vector $\bm{z}$) denotes the element-wise exponential, $\imagUnit$ is the imaginary unit, and $\varphi^{(l)}$ is the angle of an object located at $\bm{\theta} = [x\;\;y]^\top$ relative to $\varphi_{\mathrm{R}}^{(l)}$.
    The distance-dependent part is $\bm{\psi}_{\mathrm{d}}\bigl(d^{(l)}\bigr) = \frac{\lambda_\text{c}}{(4\pi)^{3/2}( d^{(l)})^2}e^{-\imagUnit\,2\pi\,(2d^{(l)}/c)\,\bm{f}}$, where $c$ is the speed of light and $d^{(l)} = \|\bm{\theta} - \bm{\theta}_{\mathrm{R}}^{(l)}\|$ is the distance from the $l$th radar to $\bm{\theta}$, with $\|\cdot\|$ denoting the Euclidean norm.
        \footnote{With this definition, $|\alpha_k^{(l)}|^2$ corresponds the radar cross section of the $k$th object in direction of the $l$th radar, i.e., the unit of $\alpha_k^{(l)}$ is meters.}
    The vector $\bm{f} = \bigl[\tfrac{-N_{\mathrm{f}} + 1}{2}\Delta_{\mathrm{f}} \ist\iist \tfrac{-N_{\mathrm{f}} + 2}{2}\Delta_{\mathrm{f}}  \ist\iist \cdots  \ist\iist \tfrac{N_{\mathrm{f}} - 1}{2}\Delta_{\mathrm{f}}\bigr]^\transpose$ specifies the equally spaced baseband frequency points at which the signal is sampled, with spacing $\Delta_{\mathrm{f}}$.
        \footnote{The definition of $\bm{f}$ given above is used for odd $N_{\mathrm{f}}$, whereas for even $N_{\mathrm{f}}$ we use $\bm{f}=[\frac{-N_{\mathrm{f}}}{2}\Delta_{\mathrm{f}} \ist\iist \frac{-N_{\mathrm{f}}+1}{2}\Delta_{\mathrm{f}} \ist\cdots\ist \frac{N_{\mathrm{f}}-1}{2}\Delta_{\mathrm{f}}]^\transpose$.}
    
    Note that the presented method can be readily generalized---e.g., to other array geometries, wideband models, or completely different sensor modalities---by appropriately choosing $\bm{\psi}^{(l)}(\bm{\theta})$ in \eqref{eq:signal-model-sum}.

	\section{Probabilistic Model and Sparse Bayesian Learning}
	\label{sec:sbl}

	To estimate the number of objects $K$, we transform the problem into a sparse signal reconstruction problem such that $K$ is estimated as the number of nonzero components of a sparse vector.
	Let $K_{\text{max}}\geq K$ be the largest possible number of objects (e.g.\ $K_{\text{max}}=N$) and $\alpha_k=0$ for $K<k \leq K_{\text{max}}$ we rewrite \eqref{eq:signal-model-sum} as a sparse linear system
	\begin{equation}
		\bm{y}^{(l)} = \bm{\Psi}^{(l)}(\bm{\Theta})\bm{\alpha}^{(l)} + \bm{v}^{(l)}
	\end{equation}
	where $\bm{\Theta}=\big[\bm{\theta}_1 \ist \iist \bm{\theta}_2 \ist \cdots \ist \bm{\theta}_{K_{\text{max}}} \big]\in \mathbb{R}^{2\times K_{\text{max}}}$ is the joint parameter matrix with each of the $K_{\text{max}}$ columns of $\bm{\Theta}$ corresponding to the XY-coordinates of a single (potential) object, $\bm{\Psi}^{(l)}(\bm{\Theta}) = \big[\bm{\psi}^{(l)}(\bm{\theta}_1)\ist\iist \bm{\psi}^{(l)}(\bm{\theta}_2) \ist\cdots\ist \bm{\psi}^{(l)}(\bm{\theta}_M)\big]$ is a dictionary matrix consisting of columns $\bm{\psi}^{(l)}(\bm{\theta}_k)$, $k=1,\dots,K_{\text{max}}$, each of which is parameterized by the position $\bm{\theta}_k$ of a single object, and $\bm{\alpha}^{(l)}=[\alpha_1 \ist\iist \alpha_2 \ist \cdots \alpha_{K_{\text{max}}}]^\transpose$ is a sparse vector of complex object reflectivities. Specifically, $\bm{\alpha}^{(l)}=[\alpha_1^{(l)} \ist\iist \alpha_2^{(l)} \ist \cdots \ist \alpha_K^{(l)} \ist\iist 0 \ist \cdots \ist 0]^\transpose$, $l=1,\dots,L$ are vectors with $K$ nonzero elements corresponding to the reflectivities of the $K$ actual objects, and $K_{\text{max}}-K$ zeros.
    The likelihood of observing $\bm{y}^{(l)}$ given $\bm{\Theta}$ and $\bm{\alpha}^{(l)}$ is
    \begin{equation}\label{eq:likelihood}
		p(\bm{y}^{(l)}|\bm{\alpha}^{(l)},\bm{\bm{\Theta}},\lambda^{(l)}) = \mathrm{CN}(\bm{y}^{(l)};\, \bm{\Psi}^{(l)}(\bm{\Theta})\bm{\alpha}^{(l)},\, (\lambda^{(l)}\bm{\Lambda}_{\text{v}}^{(l)})^{-1})
    \end{equation}
    where $\mathrm{CN}(\bm{x};\,\bm{\mu},\,\bm{\Sigma})=|\pi \bm{\Sigma}|^{-1} e^{-(\bm{x}-\bm{\mu})^\hermitian \bm{\Sigma}^{-1}(\bm{x}-\bm{\mu})}$ denotes the \gls{pdf} of multivariate complex Gaussian random variable $\bm{x}$ with mean $\bm{\mu}$ and covariance $\bm{\Sigma}$, and $(\cdot)^\hermitian$ denotes the hermitian transpose.\footnote{Note that the proposed method can readily be used for (non-complex) Gaussian models \cite{Faul2001}.}
    Assuming that the noise $\bm{v}^{(l)}$ is independent across all radars $l=1,\dots,L$ yields the joint likelihood
    \begin{align}
        p(\bm{Y}|\bm{A},\bm{\Theta},\bm{\lambda}) = \prod_{l=1}^{L} p(\bm{y}^{(l)}|\bm{\alpha}^{(l)},\bm{\bm{\Theta}},\lambda^{(l)})
    \end{align}
    where $\bm{Y}=[\bm{y}^{(1)} \ist\iist \bm{y}^{(2)} \ist\cdots\ist \bm{y}^{(L)}]$ is the matrix of all observations $\bm{y}^{(l)}$, $\bm{A}=[\bm{\alpha}^{(1)} \ist\iist \bm{\alpha}^{(2)} \ist\cdots\ist \bm{\alpha}^{(L)}]\in \mathbb{C}^{K_{\text{max}}\times L}$ is the matrix of all amplitudes, consisting of $K$ nonzero rows and $K_{\text{max}}-K$ rows of zeros, and $\bm{\lambda} = [\lambda^{(1)}\ist\iist \lambda^{(2)}\ist\cdots\ist \lambda^{(L)}]^\transpose$.

    \subsection{Sparse Bayesian Learning}
    
    To jointly estimate the number of objects $K$ and their positions $\bm{\theta}_k$, $k=1,\dots,K$, the proposed multidictionary \gls{sbl}-based method combines the fast update rule \cite{Faul2001,shutin2011TSP:fastVSBL,MoePerWitLei:Arxiv2023BSBL}, continuous estimation of dictionary parameters \cite{hansen2014SAM:SBL, hansen2018TSP:SuperFastLSE, GreLeiWitFle:TWC2024, MoePerWitLei:TSP2024}, and block-sparse models \cite{babacan2014TSP:variationalBSBL,MoePerWitLei:Arxiv2023BSBL}. To obtain a row-sparse sparse estimate of $\bm{A}$, (i.e., sparse estimates of $\bm{\alpha}^{(l)}$, $l=1,\dots,L$, with shared sparsity patterns), we introduce an improper prior with density \cite{palmerNIPS2003, MoePerWitLei:Arxiv2023BSBL}
	\begin{align}\label{eq:prior-convex} 
		p(\bm{A})
        \rmv\rmv &= \rmv\rmv
        \prod_{k=1}^{K_{\text{max}}}\sup_{\gamma_k > 0}\bigg(\prod_{l=1}^{L}\mathrm{CN}(\alpha_k^{(l)};0, \gamma_k^{-1}) \bigg) 
        \rmv\rmv \propto \rmv\rmv
        \prod_{k=1}^{K}\frac{1}{(\|\bm{A}{[k, \cdot]}\|^{2})^{L}}
	\end{align}
    that is independent indentically distributed across $k=1,\dots,K_{\text{max}}$, where $\bm{A}{[k,\cdot]}$ denotes the $k$th row of $\bm{A}$.
    By omitting the maximization over the hyperparameters $\bm{\gamma}=[\gamma_1\ist\iist \gamma_2\ist\cdots\ist\gamma_{K_{\text{max}}}]^\transpose$ in \eqref{eq:prior-convex}, we obtain a lower bound
    \begin{equation}        
        \tilde{p}(\bm{A};\bm{\gamma}) = \prod_{l=1}^{L} \mathrm{CN}(\bm{\alpha}^{(l)};\,0,\,\bm{\Gamma}^{-1})\leq p(\bm{A})    
    \end{equation}
    parameterized by $\bm{\gamma}$, where $\bm{\Gamma}=\diag(\bm{\gamma})$ is diagonal matrix with the elements of $\bm{\gamma}$ along its main diagonal.
    \Gls{sbl} then proceeds to estimate $\bm{\Theta}$, $\bm{\gamma}$ and $\bm{\lambda}$ by maximizing a lower bound $\mathcal{L}(\bm{\Theta},\bm{\gamma},\bm{\lambda})$ on the (log) marginal likelihood $p(\bm{Y}|\bm{\Theta},\bm{\lambda})$ of $\bm{Y}$ given $\bm{\Theta}$ and $\bm{\lambda}$ \cite{palmerNIPS2003,wipf2011TIP}
    \vspace{-1mm}
    \begin{align}
		(\hat{\bm{\Theta}},\hat{\bm{\gamma}},\hat{\bm{\lambda}})=\underset{{\bm{\Theta},\bm{\gamma}\succeq \bm{0},\bm{\lambda}}}{\arg\max}\  \mathcal{L}(\bm{\Theta},\bm{\gamma},\bm{\lambda})\\[-7mm]\nn
    \end{align}
    where
     \vspace{-1mm}
    \begin{align}
		\mathcal{L}(\bm{\Theta},\bm{\gamma},\bm{\lambda}) &= \log \int p(\bm{Y}|\bm{A},\bm{\Theta},\bm{\lambda})\tilde{p}(\bm{A};\bm{\gamma})\,\mathrm{d}\bm{A} \nn \\
        &
        \label{eq:sbl-objective}
        = \sum_{l=1}^{L} -(\bm{y}^{(l)})^\hermitian \big(\bm{C}^{(l)}\big)^{-1}\bm{y}^{(l)} - \log |\bm{C}^{(l)}| + \text{const.}\\[-7mm]\nn
    \end{align}
    and $\bm{C}^{(l)} = \bm{\Psi}^{(l)}(\bm{\Theta})\bm{\Gamma}^{-1}\bm{\Psi}^{(l)}(\bm{\Theta})^\hermitian + (\lambda^{(l)} \bm{\Lambda}_{\text{v}}^{(l)})^{-1}$. That is, the multi-dictionary \gls{sbl} objective \eqref{eq:sbl-objective} is a sum over classic \gls{sbl} objective functions \cite[Eq. (8)]{Faul2001} with each term corresponding to the observations of a single sensor.
	
	Once estimates $\hat{\bm{\gamma}}$, $\hat{\bm{\Theta}}$, and $\hat{\bm{\lambda}}$ are obtained, the (approximate) posterior densities of $\bm{\alpha}^{(l)}$ conditional on $\bm{y}^{(l)}$, $\hat{\bm{\gamma}}$, $\hat{\bm{\Theta}}$, and $\hat{\bm{\lambda}}$ are independent across $l=1,\dots,L$, i.e., $p(\bm{A}|\bm{Y},\hat{\bm{\Theta}},\hat{\bm{\lambda}};\hat{\bm{\gamma}}) = \prod_{l=1}^{L} p(\bm{\alpha}^{(l)}|\bm{y}^{(l)},\hat{\bm{\Theta}},\hat{\lambda}^{(l)};\hat{\bm{\gamma}})$, where
    \begin{align}
        p(\bm{\alpha}^{(l)}|\bm{y}^{(l)},\hat{\bm{\Theta}},\hat{\lambda}^{(l)};\hat{\bm{\gamma}}) &= \mathrm{CN}(\bm{\alpha}^{(l)};\,\hat{\bm{\alpha}}^{(l)},\, (\hat{\bm{\Lambda}}_{\bm{\alpha}}^{(l)})^{-1})
    \end{align}
    with
	\begin{align}
        \label{eq:amplitudes-mean}
		\hat{\bm{\alpha}}^{(l)} &= (\hat{\bm{\Lambda}}_{\bm{\alpha}}^{(l)})^{-1}(\hat{\bm{\Psi}}^{(l)})^\hermitian\hat{\bm{\Lambda}}_{\text{v}}^{(l)}\bm{y}^{(l)} \\ 
        \label{eq:amplitudes-precision}
		\hat{\bm{\Lambda}}_{\bm{\alpha}}^{(l)} &= (\hat{\bm{\Psi}}^{(l)})^\hermitian \hat{\bm{\Lambda}}_{\text{v}}^{(l)}\hat{\bm{\Psi}}^{(l)} + \hat{\bm{\Gamma}}
	\end{align}
	$\hat{\bm{\Lambda}}_{\text{v}}^{(l)}=\hat{\lambda}^{(l)}\bm{\Lambda}_{\text{v}}^{(l)}$, $\hat{\bm{\Psi}}^{(l)}=\bm{\Psi}^{(l)}(\hat{\bm{\Theta}})$, and $\hat{\bm{\Gamma}}=\diag(\hat{\bm{\gamma}})$.
	This approach is known to result in many estimates $\hat{\gamma}_k$ to diverge to infinity and so do the corresponding elements on the main diagonal of $\hat{\bm{\Lambda}}_{\bm{\alpha}}^{(l)}$. Thus, $p(\bm{\alpha}^{(l)}|\bm{y}^{(l)},\hat{\bm{\Theta}}, \hat{\lambda}^{(l)};\hat{\bm{\gamma}})$ collapses to a Dirac-delta in many dimension resulting in a amplitude vectors $\bm{\alpha}^{(l)}$ with many elements having value of zero with probability one, i.e., sparse vectors.
    Furthermore, all $\bm{\alpha}^{(l)}$ are informed by the same hyperparameters $\bm{\gamma}$ and, thus, share the same sparsity pattern. 
	
	\subsection{Object Parameter Updates}
    \label{sec:sbl:mmv-updates}
    Finding the global maximum of $\mathcal{L}(\bm{\Theta},\bm{\gamma},\bm{\lambda})$ with respect to all parameters $(\bm{\Theta},\bm{\gamma},\bm{\lambda})$ jointly is computationally prohibitive.
    Thus, we maximize $\mathcal{L}(\bm{\Theta},\bm{\gamma},\bm{\lambda})$ using coordinate ascent with respect to the tuple of parameters $(\bm{\theta}_k,\gamma_k)$ corresponding to a single object at a time, while keeping the remaining parameters fixed.
    Let $\bm{M}_{\sim k}^{(l)} = \bm{I}-\hat{\bm{\Psi}}_{\sim k}^{(l)} ((\hat{\bm{\Psi}}_{\sim k}^{(l)})^\hermitian\hat{\bm{\Lambda}}_{\text{v}}^{(l)}\hat{\bm{\Psi}}_{\sim k}^{(l)} + \hat{\bm{\Gamma}}_{\sim k})^{-1}(\hat{\bm{\Psi}}_{\sim k}^{(l)})^\hermitian\hat{\bm{\Lambda}}_{\text{v}}^{(l)}$ and $\hat{\bm{\Psi}}_{\sim k}^{(l)}=\bm{\Psi}^{(l)}([\hat{\bm{\theta}}_1 \ist\iist \hat{\bm{\theta}}_2 \ist\cdots\ist \hat{\bm{\theta}}_{k-1} \ist\iist\hat{\bm{\theta}}_{k+1} \ist\cdots\ist \hat{\bm{\theta}}_{K_{\text{max}}}])$ denote a matrix of all but the $k$th column of the dictionary $\hat{\bm{\Psi}}^{(l)}$, and $\hat{\bm{\Gamma}}_{\sim k}=\diag(\hat{\bm{\gamma}}_{\sim k})$ with $\hat{\bm{\gamma}}_{\sim k}$ as the vector $\hat{\bm{\gamma}}$ with the $k$th element removed.
    Following the derivation of \cite{Faul2001}, 
    the dependency of $\mathcal{L}(\bm{\Theta}, \bm{\gamma},\bm{\lambda})$ on the tuple $(\bm{\theta}_k,\gamma_k)$ corresponding to the $k$th object can be made explicit as $\mathcal{L}(\bm{\Theta},\bm{\gamma},\bm{\lambda})=\mathcal{L}_{\sim k} + \ell_k(\bm{\theta}_k,\gamma_k)$ where $\mathcal{L}_{\sim k}$ is some function that does not depend on $(\bm{\theta}_k,\gamma_k)$, and
    \begin{equation} \label{eq:partial-likelihood}
        \ell_k(\bm{\theta}_k,\gamma_k) = \sum_{l=1}^{L} \frac{|\mu_k^{(l)}(\bm{\theta}_k)|^2/s_k^{(l)}(\bm{\theta}_k)}{1+\gamma_k s_k^{(l)}(\bm{\theta}_k)} + \log \frac{\gamma_k s_k^{(l)}(\bm{\theta}_k)}{1+\gamma_k s_k^{(l)}(\bm{\theta}_k)}
    \end{equation}
    with
    \begin{align}
		\label{eq:sk}
		s_k^{(l)}(\bm{\theta}_k) = \big(\bm{\psi}^{(l)}(\bm{\theta}_k)^\hermitian \, \hat{\bm{\Lambda}}_{\text{v}}^{(l)} \,\bm{M}_{\sim k}^{(l)}\bm{\psi}^{(l)}(\bm{\theta}_k)\big)^{-1} \\
		\label{eq:qk}
		  \mu_k^{(l)}(\bm{\theta}_k) = s_k^{(l)}(\bm{\theta}_k) \bm{\psi}^{(l)}(\bm{\theta}_k)^\hermitian \, \hat{\bm{\Lambda}}_{\text{v}}^{(l)} \, \bm{M}_{\sim k}^{(l)}\bm{y}^{(l)}
          \,.
	\end{align}

    Since there is no analytic solution to the joint maximum of $\ell_k(\bm{\theta}_k,\gamma_k)$ we resort to coordinate ascent.
    First, we find
    \begin{equation}
        \hat{\bm{\theta}}_k=\arg\max_{\bm{\theta}}\ell_k(\bm{\theta},\hat{\gamma}_k)
    \end{equation}
    by means of a numeric optimizer using the previous estimate $\hat{\gamma}_k$.
    Next, we update $\hat{\gamma}_k$.
    Taking the derivative of $\ell_k(\hat{\bm{\theta}},\gamma_k)$ with respect to $\gamma_k$ yields the fixed point equation
    \begin{align} \label{eq:fixed-point-mmv}
        0=\sum_{l=1}^{L}\frac{1-\gamma_k(|\mu_k^{(l)}(\hat{\bm{\theta}}_k)|^2-s_k^{(l)}(\hat{\bm{\theta}}_k))}{(1+\gamma_k s_k^{(l)}(\hat{\bm{\theta}}_k))^2}
    \end{align}
    which can be solved by finding the solutions to the polynomial equation $P_k(\gamma) = 0$ (i.e., by finding the roots of $P_k$), where
    \begin{equation} \label{eq:fixed-point-polynomial}
        P_k(\gamma) \rmv\rmv:=\rmv\rmv \sum_{l=1}^{L}\big(1-\gamma(|\mu_k^{(l)}(\hat{\bm{\theta}}_k)|^2-s_k^{(l)}(\hat{\bm{\theta}}_k)\big) \hspace*{-2mm} \prod_{j=1,j \neq l}^{L}\hspace*{-2mm} (1+\gamma s_k^{(j)}(\hat{\bm{\theta}}_k))^2
    \end{equation}
    is a polynomial in $\gamma$ of degree $2L-1$.
    Let $\mathcal{G}_k :=\{\gamma > 0:P_k(\gamma)=0\}$ denote the set of positive, real-valued roots of $P_k$, 
    we proceed to update $\hat{\gamma}_k$ using
    \begin{equation}\label{eq:gamma-update-mmv}
        \hat{\gamma}_k = \begin{cases}
            \arg \max_{\gamma\in\mathcal{G}_k} \ell_k(\hat{\bm{\theta}}_k,\gamma) & \text{if } \mathcal{G}_k \neq \emptyset \\
            \infty & \text{else}
        \end{cases}
    \end{equation}
    where $\emptyset$ denotes the empty set, i.e., we set $\hat{\gamma}_k$ to the solution of \eqref{eq:fixed-point-mmv} that increases $\ell_k(\hat{\bm{\theta}}_k,\cdot)$ the most.

    \subsubsection*{Single Radar}
    Lets consider the single-radar case (i.e., $L=1$) in this paragraph, where we omit the superscript $(\cdot)^{(l)}$ for ease of notation.
    In this case, \eqref{eq:fixed-point-mmv} has a single positive solution if $|\mu_k(\bm{\theta}_k)|^2>s_k(\bm{\theta}_k)$, whereas no positive solution exists for otherwise \cite{Faul2001}.
    Hence, for any given location $\bm{\theta}$, $\ell_l(\bm{\theta},\cdot)$ achieves its maximum at
    \begin{equation}
        \gamma_k^\star(\bm{\theta}) = \begin{cases}
            \frac{1}{|\mu_k(\bm{\theta}_k)|^2-s_k(\bm{\theta}_k)} & \text{if } Q_k(\bm{\theta}_k) > 1 \\
            \infty & \text{else}
        \end{cases}
    \end{equation}
    where $Q_k(\bm{\theta}):=|\mu_k(\bm{\theta})|^2/s_k(\bm{\theta})$ can be recognized as the component \gls{snr} \cite{leitinger2020Asilomar}.
    Inserting the optimal value $\gamma_k^\star$ into $\ell_k$ yields $	\ell_k^\star(\bm{\theta}) := \ell_k\big(\bm{\theta},\gamma_k^\star(\bm{\theta})\big)$, where
    \begin{align} \label{eq:concentrated-partial-likelihood}
		\ell_k^\star(\bm{\theta}) &= \begin{cases}
			Q_k(\bm{\theta}) \rmv\rmv - \rmv\rmv 1 \rmv\rmv - \rmv\rmv \log Q_k(\bm{\theta}) & \text{if }Q_k(\bm{\theta}) > 1 \\
			0 & \text{else}
		\end{cases}
        \, .
	\end{align}
    It is straightforward to show, that $\ell_k^\star(\bm{\theta})$ is an increasing function of the component \gls{snr} $Q_k$. Hence, in the single radar case we find the supremum of $\ell_k(\bm{\theta}_k,\gamma_k)$ jointly as
    \begin{align}
        \hat{\bm{\theta}}_k &= \arg\max_{\bm{\theta}}Q_k(\bm{\theta}) \\
        \hat{\gamma}_k &= \gamma_k^\star(\hat{\bm{\theta}}_k) 
    \end{align}
    i.e., \gls{sbl} objective $\mathcal{L}(\bm{\Theta},\bm{\gamma},\bm{\lambda})$ is maximized with respect to the parameters of a single object $(\bm{\theta}_k,\gamma_k)$ by the position $\bm{\theta}$ that maximizes the component \gls{snr} $Q_k(\bm{\theta})$ and the object is estimated to have nonzero amplitude if, and only if, the component \gls{snr} exceeds unity.

    \emph{Identical Sensors:}
    Lets assume that the array responses $\bm{\psi}_k^{(l)}=\bm{\psi}_k$ and noise distribution (i.e., $\lambda^{(l)}\bm{\Lambda}_{\text{v}}^{(l)}$) are identical across all $l=1,\dots,L$, corresponding to, e.g., multiple measurement vectors obtained from the same radar.
    In this case, the variables $s_k^{(l)}=s_k$ defined in \eqref{eq:sk} do not depend on $l$ since all $\bm{\psi}^{(l)}$ are equal such that \eqref{eq:fixed-point-mmv} can be simplified to
    \begin{equation}
        0=\frac{1-\gamma_k(\bar{\mu}_k^2(\hat{\bm{\theta}}_k)-s_k(\hat{\bm{\theta}}_k))}{(1+\gamma s_k(\hat{\bm{\theta}}_k))^2}
    \end{equation}
    where $\bar\mu_k^2(\hat{\bm{\theta}}_k) = 1/L \cdot \sum_{l=1}^{L}|\mu_k^{(l)}(\hat{\bm{\theta}}_k)|^2$.
    Hence, the multiple measurement vector case with identical array responses can be treated same as the single measurement vector case  $L=1$, except that $|\mu_k(\hat{\bm{\theta}}_k)|^2$ is replaced by $\bar{\mu}_k^2(\hat{\bm{\theta}}_k)$.
    
    \subsubsection*{Additional Thresholding}
    It is experimentally known, that the updates for the object parameters $(\bm{\theta}_k, \gamma_k)$, $k=1,\dots,K$ (particular the update of the hyperparameter $\gamma_k$ governing object existence) described in this section results in a relatively high number of false alarms \cite{leitinger2020Asilomar}.
    To reduce the number of false alarms, we require all detections to exceed a certain component \gls{snr} (averaged over $l=1,\dots,L$). Specifically, when updating the value of $\hat{\gamma}_k$ we use
    \begin{equation}\label{eq:gamma-update-mmv-threshold}
        \hat{\gamma}_k = \begin{cases}
            \arg \max_{\gamma\in\mathcal{G}_k} \ell_k(\hat{\bm{\theta}}_k,\gamma) & \text{if } \mathcal{G}_k \neq \emptyset \text{ and } \bar{Q}_k(\hat{\bm{\theta}}_k)>\chi \\
            \infty & \text{else}
        \end{cases}
    \end{equation}
    instead of \eqref{eq:gamma-update-mmv},
    where $\bar{Q}_k(\bm{\theta})=1/L \cdot \sum_{l=1}^{L}|\mu_k^{(l)}(\bm{\theta})|^2/s_k^{(l)}(\bm{\theta})$ is the component \gls{snr} averaged over all radars and $\chi\geq 1$ is the minimum required average component \gls{snr}.
    See \cite{leitinger2020Asilomar} for a detailed analysis of the relation between the threshold $\chi$ and the false alarm rate for the single radar case.

	\subsection{Noise Power Estimate}
	\label{sec:sbl:noise-update}

	No closed form solution for the update of the noise parameters $\lambda^{(l)}$ is available.
    Thus, we resort to an \gls{em} update that increases a lower bound $\mathcal{L}^{\text{EM}}(\bm{\lambda},\hat{\bm{\lambda}}^{\text{old}})\leq \mathcal{L}(\hat{\bm{\Theta}},\hat{\bm{\gamma}},\bm{\lambda})$ given the current estimates $\hat{\bm{\lambda}}^{\text{old}}$, $\hat{\bm{\Theta}}$ and $\hat{\bm{\gamma}}$ \cite{Bishop2006, tzikas2008:VAEM}.
    We consider $\bm{Y}$ as the observed data and $(\bm{Y},\bm{A})$ as the complete data.
    An improved estimate $\hat{\bm{\lambda}}^{\text{new}} = \arg\max_{\bm{\lambda}} \mathcal{L}^{\text{EM}}(\bm{\lambda},\hat{\bm{\lambda}}^{\text{old}})$ is obtained as maximizer of the \gls{em} objective
	\begin{align} \label{eq:noise-em-objective}
		\mathcal{L}^{\text{EM}}(\bm{\lambda},\hat{\bm{\lambda}}^{\text{old}}) &= \int p(\bm{A}|\bm{Y},\hat{\bm{\Theta}},\hat{\bm{\lambda}}^{\text{old}};\hat{\bm{\gamma}}) \nn\\
      & \hspace{10mm}\times 
      \log p(\bm{Y},\bm{A}|\hat{\bm{\Theta}},\bm{\lambda};\hat{\bm{\gamma}}) \,\mathrm{d}\bm{A}
	\end{align}
	where $p(\bm{Y},\bm{A}|\hat{\bm{\Theta}},\bm{\lambda};\hat{\bm{\gamma}}) = p(\bm{Y}|\bm{A},\hat{\bm{\Theta}},\bm{\lambda})p(\bm{A};\hat{\bm{\gamma}})$.
	Solving the integral and setting the derivative of $\mathcal{L}^{\text{EM}}$ to zero yields the estimates
	\begin{align}
		\hat{\lambda}^{(l),\text{new}}
        &=\frac{N}{(\hat{\bm{y}}_{\text{res}}^{(l)})^\hermitian\bm{\Lambda}_{\text{v}}^{(l)}\hat{\bm{y}}_{\text{res}}^{(l)} + \trace\big((\hat{\bm{\Lambda}}_{\bm{\alpha}}^{(l)})^{-1}(\hat{\bm{\Psi}}^{(l)})^\hermitian \bm{\Lambda}_{\text{v}}^{(l)}\hat{\bm{\Psi}}^{(l)}\big)}
		\label{eq:lambda-update}
	\end{align}
    for $l=1,\dots,L$,
    where $\hat{\bm{y}}_{\text{res}}^{(l)} = \bm{y}^{(l)}-\hat{\bm{\Psi}}^{(l)}\hat{\bm{\alpha}}^{(l)}$, $\trace(\cdot)$ is the trace operator, and $\hat{\bm{\alpha}}^{(l)}$ and $\hat{\bm{\Lambda}}_{\bm{\alpha}}^{(l)}$ are the mean and precision, respectively, of $p(\bm{\alpha}^{(l)}|\bm{y}^{(l)},\hat{\bm{\Theta}},\hat{\lambda}^{(l),\text{old}};\hat{\bm{\gamma}})$ given by \eqref{eq:amplitudes-mean} and \eqref{eq:amplitudes-precision}, respectively.

	The updates given in subsections~\ref{sec:sbl:mmv-updates} and \ref{sec:sbl:noise-update} are repeated in a round-robin fashion until convergence or until a fixed number of iterations is exceeded.

	\section{Algorithm}
	\label{sec:algorithm}

	We define an algorithm by combining the coordinate-ascent updates of estimates $(\hat{\bm{\theta}}_k,\hat{\gamma}_k)$ together with an initialization and update-schedule as summarized in Algorithm\ref{alg:sbl}.
        \footnote{Python code is available at \url{https://doi.org/10.3217/kcn0n-03509}}

    We start with an empty model $\hat{\gamma}_k=\infty$, $k=1,\dots,K_{\text{max}}$, and initialize the noise parameters as $\hat{\lambda}^{(l)} = 10 \cdot N /(\bm{y}^{(l)})^\hermitian\bm{\Lambda}_{\text{v}}^{(l)}\bm{y}^{(l)}$.
    In each iteration, we first go through all ``active'' objects currently in the model $k\in \{1\leq k \leq K_{\text{max}} : \hat{\gamma}_k<\infty\}$ and update their estimated positions $\hat{\bm{\theta}}_k$. To do so, we use a numeric optimizer (Python's \verb|scipy.optimize|) to maximize
    \begin{align} \label{eq:update-theta}
        \hat{\bm{\theta}}_k = \arg\max_{\bm{\theta}}\begin{cases}
            Q_k(\bm{\theta}) & \text{if } L=1 \\
            \ell_k(\bm{\theta},\hat{\gamma}_k) & \text{if } L>1
        \end{cases}
        \, .
    \end{align}
	Once $\hat{\bm{\theta}}_k$ is obtained, we also update $\hat{\gamma}_k$ using \eqref{eq:gamma-update-mmv-threshold}, potentially setting it to $\hat{\gamma}_k=\infty$ and thereby ``deactivating'' $k$th the object.
	After iterating through all currently existing objects, we search for a new object to detect.
    To do so, we chose any $k$ corresponding to a ``deactivated'' object (i.e., $\hat{\gamma}_k=\infty$) and find the position of this potential object by finding the position $\hat{\bm{\theta}}_{k}$ that maximizes the average component \gls{snr} $\bar{Q}_k(\bm{\theta})$.
    To aid the numeric optimizer during the search for new objects, we first evaluate $\bar{Q}_{k}(\bm{\theta})$ on a sub-Nyquist grid $\bar{\bm{\Theta}}$ and initialize the numeric optimizer with the location of the maximum on the grid.
	Again, we use the computed position $\hat{\bm{\theta}}_{k}=\arg\max_{\bm{\theta}}\bar{Q}_k(\bm{\theta})$ to set $\hat{\gamma}_k$ according to \eqref{eq:gamma-update-mmv-threshold}, potentially including it as new object in the model.
    Finally, we update the estimate of the noise parameter $\hat{\lambda}$ using \eqref{eq:lambda-update}.
	Once the algorithm is converged (or a maximum number of iterations is reached), we also compute the mean and covariance, $\hat{\bm{\alpha}}^{(l)}$ and $\hat{\bm{\Lambda}}_{\bm{\alpha}}^{(l)}$ for $l=1,\dots,L$, respectively, of the (approximate) posterior $p(\bm{A}|\bm{Y},\hat{\bm{\Theta}},\hat{\bm{\lambda}};\hat{\bm{\gamma}})= \prod_{l=1}^{L}p(\bm{\alpha}^{(l)}|\bm{y}^{(l)},\hat{\bm{\Theta}},\hat{\lambda}^{(l)};\hat{\bm{\gamma}})$.	
	
	    \begin{algorithm}[tbp]
		\renewcommand{\algorithmicrequire}{\textbf{Input:}}
		\renewcommand{\algorithmicensure}{\textbf{Output:}}
		\caption{SBL for Multiple Parameterized Dictionaries}
		\label{alg:sbl}
		\begin{algorithmic}
            \REQUIRE Signals $\bm{y}^{(l)}$, $l=1,\dots,L$, noise precision $\bm{\Lambda}_{\text{v}}^{(l)}$, array responses $\bm{\psi}^{(l)}: \mathbb{R}^2 \mapsto \mathbb{C}^{N \times 1}$, threshold $\chi$ and grid $\bar{\bm{\Theta}}$.
			\ENSURE Object locations $\hat{\bm{\Theta}}$, hyperparameters $\hat{\bm{\gamma}}$, amplitudes $\hat{\bm{\alpha}}^{(l)}$, $l=1,\dots,L$, and noise parameter $\hat{\lambda}$.
			
			\STATE Initialize $\hat{\lambda}^{(l)} \leftarrow 10 \cdot N /((\bm{y}^{(l)})^\hermitian\bm{\Lambda}_{\text{v}}^{(l)}\bm{y}^{(l)})$, and $\hat{\gamma}_k \leftarrow \infty$ for $l=1,\dots,L$, and $k=1,\dots,K_{\text{max}}$, respectively.%
			
			\WHILE{not converged}
			\FOR{$k\in \{1\leq k\leq K_{\text{max}} : \hat{\gamma}_k<\infty\}$}
                    \STATE Update $\hat{\bm{\theta}}_k$ using \eqref{eq:update-theta}.
       		\STATE Update $\hat{\gamma}_k$ using \eqref{eq:gamma-update-mmv-threshold}.
			\ENDFOR	
			\STATE $k\leftarrow$ chose any from $\{1\leq k\leq K_{\text{max}} : \hat{\gamma}_k=\infty\}$.
            \STATE $\hat{\bm{\theta}}_k \leftarrow \arg\max_{\bm{\theta}} \bar{Q}_k(\bm{\theta})$.
            \STATE Update $\hat{\gamma}_k$ using \eqref{eq:gamma-update-mmv-threshold}.
			\STATE Update $\hat{\lambda}^{(l)}$ for $l=1,\dots,L$ using \eqref{eq:lambda-update}.
			\ENDWHILE
            \STATE Calculate $\hat{\bm{\alpha}}^{(l)}$ and $\hat{\bm{\Lambda}}_{\bm{\alpha}}^{(l)}$ for $l=1,\dots,L$ using \eqref{eq:amplitudes-mean} and \eqref{eq:amplitudes-precision}, respectively.
		\end{algorithmic}
	\end{algorithm}

	Note that the quantities $\mu_k(\bm{\theta})$ and $s_k(\bm{\theta})$ can be computed in an efficient manner by considering the symmetries of the matrix
	\begin{equation}
		\hat{\bm{\Lambda}}_{\text{v}}\hat{\bm{M}}_{\sim k} = 	\hat{\bm{\Lambda}}_{\text{v}}-\hat{\bm{\Lambda}}_{\text{v}}\hat{\bm{\Psi}}_{\sim k} (\hat{\bm{\Psi}}_{\sim k}^\hermitian \hat{\bm{\Lambda}}_{\text{v}} \hat{\bm{\Psi}}_{\sim k} + \hat{\bm{\Gamma}}_{\sim k})^{-1} \hat{\bm{\Psi}}_{\sim k}^\hermitian\hat{\bm{\Lambda}}_{\text{v}}
	\end{equation}
    where we omitted the superscript $(\cdot)^{(l)}$ for ease of notation.
	Let $\bm{L}\bm{L}^\hermitian = \hat{\bm{\Lambda}}_{\text{v}}$ and $\bm{D}\bm{D}^\hermitian = \hat{\bm{\Lambda}}_{\bm{\alpha}}$ be the choleskey decompositions of $\hat{\bm{\Lambda}}_{\text{v}}$, and $\hat{\bm{\Lambda}}_{\bm{\alpha}}$, respectively, and $\bm{R} = \hat{\bm{\Lambda}}_{\text{v}} \hat{\bm{\Psi}}_{\sim k} \bm{D}^{-1}$, such that $\hat{\bm{\Lambda}}_{\text{v}}\hat{\bm{M}}_{\sim k} = \bm{L}\bm{L}^\hermitian - \bm{R}\bm{R}^\hermitian$.
	Thus we can compute
    \begin{align}
        s_k(\bm{\theta}) &= 1/(\|\bm{L}^\hermitian\bm{\psi}(\bm{\theta})\|^2 - \|\bm{R}^\hermitian\bm{\psi}(\bm{\theta})\|_{\text{F}}^2) \\
        \mu_k(\bm{\theta}) &=s_k(\bm{\theta}) \bm{\psi}(\bm{\theta})^\hermitian\bm{y}_{\text{res}}
    \end{align}
    where $\bm{y}_{\text{res}} =\hat{\bm{\Lambda}}_{\text{v}}\bm{y}-\bm{R}\bm{R}^\hermitian\bm{y}$, and $\|\cdot\|_{\text{F}}$ denotes the Frobenius norm.
    Here $\bm{y}_{\text{res}}$, $\bm{L}$ and $\bm{R}$ do not depend on $\bm{\theta}$. Thus, they can be pre-computed for the (numeric) optimization over $\bm{\theta}$ in \eqref{eq:update-theta}.

    \section{Results}
    \label{sec:results}
    
    \subsection{Single Radar}
    \label{sec:results:single-radar}
    
    We assume as single radar ($L=1$) located at the origin (i.e, $\bm{\theta}_{\mathrm{R}}^{(1)}=[0\ist\ist 0]^\transpose$) with broadside direction $\varphi_{\mathrm{R}}^{(1)}=\pi /2$ oriented in parallel to the Y-axis.
    For ease of interpretation, we do not consider a distance-dependent path loss, since the path loss can be absorbed into $\alpha_k^{(l)}$ anyways.
    Finally, we set $\bm{v}^{(1)}$ as \gls{awgn} with unit variance, i.e., $\lambda^{(1)}=1$ and $\bm{\Lambda}_{\text{v}}^{(1)}=\bm{I}$.

    We compare the presented \gls{sbl} algorithm with the \gls{nomp} algorithm \cite{mamandipoor2016TSP:NOMP}, but instead of using the Newton updates proposed in \cite{mamandipoor2016TSP:NOMP}, we employ the same numeric optimizer that is used in our \gls{sbl} implementation (i.e., Python's \verb|scipy.optimize| function). Note that unlike the proposed \gls{sbl} algorithm, the \gls{nomp} algorithm assumes the noise covariance to be fully known, i.e., it requires all $\lambda^{(l)}$ to be known in addition to $\bm{\Lambda}_{\text{v}}^{(l)}$.
    Furthermore, with the parameters given in Section~\ref{sec:signal-model}, the radar has a resolution of roughly $7.5\,\text{m}$ in range and $16.5^\circ$ in angle (corresponding to a cross-range resolution of $8.5\,\text{m}$ at a range of $30\,\text{m}$). Thus, the performance of a \gls{cfar} detector (or similar grid-based approaches) will be significantly worse than either \gls{nomp} or the proposed \gls{sbl} approach. Similarly, we focus on \gls{sbl} as method to detect targets in individual time steps, e.g., when used as pre-processing for a multitarget tracking algorithm such as \cite{MeyerProc2018}. Thus, we consider a more thorough comparison against tracking algorithms outside the scope of this work.

    \begin{figure}
        \centering
        \includegraphics{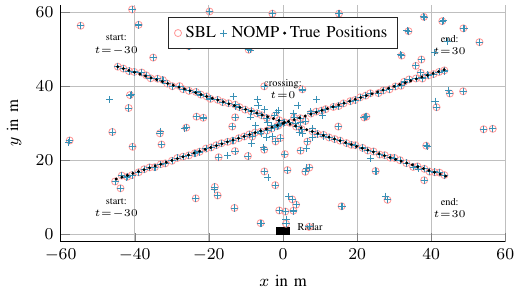}
        \caption{Tracks crossing at an angle of $36^{\circ}$ with component SNR of $20\,\text{dB}$ each, and locations of detected objects from SBL and NOMP using low thresholds ($7.4\,\text{dB}$ for SBL and $7\,\text{dB}$ for MOMP). Best viewed in color.}
        \label{fig:crossing-illustration}
    \end{figure}

    To evaluate the detection and estimation performance for both, well-separated and closely-spaced objects, we consider two objects that approach one another and cross paths, as depicted in Figure~\ref{fig:crossing-illustration}.
    Initially, these objects occupy regions where they are clearly separated in delay. At time step $t=0$ the objects are located at positions $\bm{\theta}_1 = [0,20]^\transpose$ and $\bm{\theta}_2=[0.3, 20.4]^\transpose$, i.e., separated only by $0.5\,\text{m}$.
    Note that the distance between the targets corresponds roughly to $|t|$, i.e., at $t\in\{-30,30\}$ the targets are roughly $30\,\text{m}$ apart whereas at $t=0$ the targets (almost) coincide.
    As a performance metric, we consider the \gls{ospa} \cite{schuhmacher2008TSP:OSPA} with the \gls{ospa} order set to two (Euclidean metric) and cutoff distance of $10\,\text{m}$.
    For each time step we perform $500$ Monte-Carlo runs and report the average \gls{ospa}.
    In each realization, the absolute value of the amplitudes $|\alpha_k^{(1)}|$, $k=1,2$ is chosen such that the component \gls{snr} $\|\bm{\psi}^{(1)}(\bm{\theta}_k)\alpha_k^{(1)}\|$ equals $30\,\text{dB}$ and the phase is drawn uniformly random with limits $[0, 2\pi)$.
    Both algorithms are evaluated for two different thresholds. To achieve a low false alarm rate of approximately $3.3\,\%$, the thresholds of the algorithms are set to $\chi=10\,\text{dB}$ for \gls{sbl} and, equally, a threshold of $10\,\text{dB}$ (denoted as $\tau$ in \cite{mamandipoor2016TSP:NOMP}) for \gls{nomp}.
    For a higher false alarm rate of approximately two false alarms per time step,  we set the thresholds to $\chi=7.4\,\text{dB}$ for \gls{sbl} and $7\,\text{dB}$ for \gls{nomp}.

    \begin{figure}
        \centering
        \includegraphics{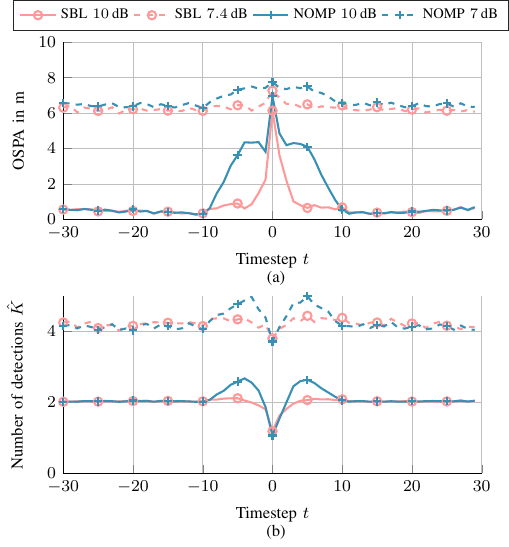}
        \caption{OSPA (a) and estimated number of objects $\hat{K}$ (b) for crossing tracks with component SNR of $30\,\text{dB}$ each. High thresholds ($10\,\text{dB}$ for both \gls{sbl} and \gls{nomp}) are shown as solid lines and low thresholds ($7.4\,\text{dB}$ for SBL $7\,\text{dB}$ for NOMP) as dashed lines.}
        \label{fig:crossing-ospa}
    \end{figure}

    Figure~\ref{fig:crossing-ospa}a shows the obtained \gls{ospa} whereas Figure~\ref{fig:crossing-ospa}b shows the estimated model order as function of the time step $t$.
    For well separated objects ($|t|>10$), the performance of both \gls{nomp} and \gls{sbl} is virtually identical in both cases.
    For medium to small separations ($5<|t|<10$), the performance of the \gls{sbl} algorithm is similar to that of $|t|>10$, whereas the number of detections (i.e., the false alarm rate) is increased for \gls{nomp}, resulting in an increased \gls{ospa} as well.
    For $|t|<5$, the probability that either of the two algorithms can accurately detect that there are two objects decreases, due the objects close proximity.
    Thus, the performance of both algorithms deteriorates, with \gls{sbl} still outperforming \gls{nomp}.
    This is in line with theoretical results concerning the advantages of Type-II Bayesian methods over Type-I Bayesian methods \cite{wipf2011TIP}.

    \subsection{Multiple Radars}
    \label{sec:results:multi-radar}
    To demonstrate the ability of the proposed multi-dictionary \gls{sbl} algorithm to fuse data from multiple radars, we consider a single object at $\bm{\theta}_1=[0\ist\iist 30]^\transpose$ with component \gls{snr} of $15\,\text{dB}$ that is observed by up to four radars located at $
    \bm{\theta}_{\text{R}}^{(l)}=[0 \ist\iist 0]^\transpose$, $[-30 \ist\iist 30]^\transpose$, $[0 \ist\iist 60]^\transpose$, and $[30 \ist\iist 30]^\transpose$ for $l=1,\dots,4$.
    Each of those radars has a $3\times 3$ \gls{mimo} array as described in Section~\ref{sec:signal-model} with broadside direction aimed towards the object and we assume \gls{awgn} with unit variance for all radars.

    \begin{figure}
    	\centering
    	\includegraphics{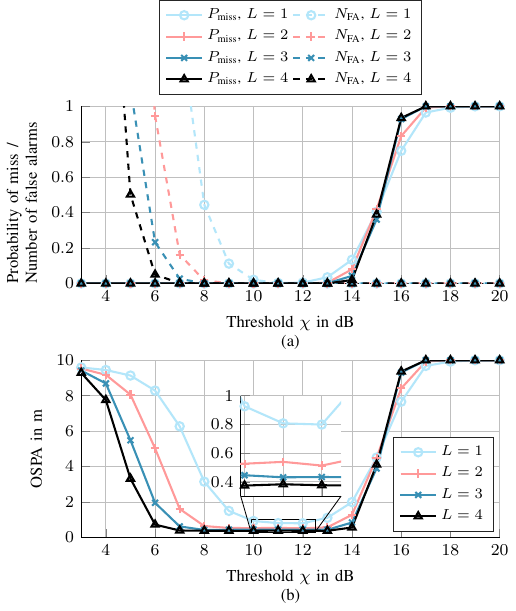}
        \caption{Detection (a) and localization (b) performance of the proposed \gls{sbl} algorithm for a single object at $\bm{\theta}_1=[0 \ist\iist 30]^\transpose$ with an component \gls{snr} of $15\,\text{dB}$ observed by multiple radars.}
        \label{fig:multiradar}
    \end{figure}

    To evaluate the detection performance as function of the threshold $\chi$ and the number of radars $L$, we consider the object to be detected if any estimated objected is located within $5\,\text{m}$ of the true object location.
    Depending on the threshold $\chi$, Figure~\ref{fig:multiradar}a shows the probability of miss detection $P_{\text{miss}}$ and the number of false alarms $N_{\text{FA}}$, i.e., the number of estimated objects not located within $5\,\text{m}$ of the true object location, averaged over $10^3$ Monte-Carlo runs.
    The number of false alarms reduces with increasing number of radars.
    This is because, when testing a position $\bm{\theta}$, the probability that the noise contribution is (jointly) high decreases with the number of (independent) observations.
    The detection performance is approximately similar for all cases.
    However, the slope of the missed detection curve is smaller for the single-radar case $L=1$ compared to $L>1$. This is again due to the noise interfering either constructively or deconstructively with the signal reflected from the object. This effect is most prominent for $L=1$ and reduces for increasing $L$ due to the noise-averaging effect of multiple observations.
    In short, increasing the number of radars helps to decrease the number of false alarms. Equivalently, this allows the use of a smaller detection threshold to increase the detection performance in low-\gls{snr} cases.

    The localization performance (in terms of the \gls{ospa}) is shown in Figure~\ref{fig:multiradar}b.
    For high thresholds ($\chi \geq 15\,\text{dB}$), the \gls{ospa} is dominated by missed detections.
    For small thresholds (depending on the number of radars $L$, e.g., $\chi < 8\,\text{dB}$ for $L=1$), the \gls{ospa} is dominated by false alarms.
    In between, e.g., for a threshold of $\chi=11\,\text{dB}$, there are almost no miss detections or false alarms (see Figure~\ref{fig:multiradar}a).
    Therefore the \gls{ospa} reflects the localization accuracy.
    It can be seen, that for $\chi=11\,\text{dB}$ an \gls{ospa} of $0.84\,\text{m}$, $0.53\,\text{m}$, $0.43\,\text{m}$, and $0.37\,\text{m}$ is achieved for $L=1,2,3,4$, respectively.

        \begin{figure}
       	\includegraphics{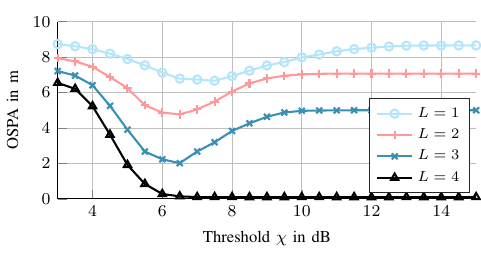}
    	\caption{OSPA of the proposed \gls{sbl} algorithm for four object observed by up to four radars when considering the path loss.}
    	\label{fig:pathloss}
    \end{figure}

    As a second multi-radar example, we simulate four objects at positions $\bm{\theta}_k=[0 \ist\iist 10]^\transpose$, $[20 \iist -\rmv\rmv 30]^\transpose$, $[0 \ist\iist 50]^\transpose$, and $[20 \ist\iist 30]^\transpose$ for $k=1,\dots,4$. All four objects are observed by up to four radars with the same positions as before. In this case, we consider the distance-dependent path loss and set the objects amplitudes $\alpha_k^{(l)}$ such that a component \gls{snr} of $30\,\text{dB}$ is achieved at a distance of $10\,\text{m}$.
    Figure~\ref{fig:pathloss} shows the \gls{ospa} achieved by the proposed algorithm for varying thresholds $\chi$.
    In this case, each additional radar improves the \gls{ospa} by a wide margin, due to the different radars achieving high \glspl{snr} in different regions of the surveilled area.

    \section{Conclusion}
    
    We introduced a multidictionary \gls{sbl} algorithm that jointly detects and estimates dictionary parameters on a continuum across multiple sensors. The algorithm has been used to detect and localize objects by fusing data from multiple \gls{mimo} radars. In the single-radar setting, the proposed multidictionary \gls{sbl} algorithm outperforms the \gls{nomp} algorithm \cite{mamandipoor2016TSP:NOMP}, especially when multiple objects are closely spaced. We illustrate how combining data from multiple radars can reduce the false alarm rate and improve localization accuracy. By choosing an appropriate parameterized dictionary, the same multi-dictionary algorithm can be adapted to other applications, e.g., direction of arrival estimation in multi-frequency ocean acoustic channels \cite{NanGemGerHidMec:SP2019}.

    \bibliographystyle{IEEEtran}
    \bibliography{IEEEabrv,sbl-radar-references}
	
\end{document}